\documentclass[11pt]{revtex4}
\usepackage{amssymb}
\usepackage{latexsym}
\usepackage{epsfig}

\usepackage{amsmath}
\begin{document}
\title{Stability of HDE model with sign-changeable interaction in Brans-Dicke theory }
\author{M. Abdollahi Zadeh$^{1}$\footnote{
m.abdollahizadeh@shirazu.ac.ir} and A.
Sheykhi$^{1,2}$\footnote{asheykhi@shirazu.ac.ir}}
\address{$^1$ Physics Department and Biruni Observatory, College of
Sciences, Shiraz University, Shiraz 71454, Iran\\
$^2$ Research Institute for Astronomy and Astrophysics of Maragha
(RIAAM), P.O. Box 55134-441, Maragha, Iran}

\begin{abstract}
We consider the Brans-Dicke (BD) theory of gravity and explore the
cosmological implications of the sign-changeable interacting
holographic dark energy (HDE) model in the background of
Friedmann-Robertson-Walker  (FRW) universe. As the system's
infrared (IR) cutoff, we choose the future event horizon, the
Granda-Oliveros (GO) and the Ricci cutoffs. For each cutoff, we
obtain the density parameter, the equation of state (EoS) and the
deceleration parameter of the system. In case of future event
horizon, we find out that the EoS parameter, $w_{D}$, can cross
the phantom line, as a result the transition from deceleration to
acceleration expansion of the universe can be achieved provided
the model parameters are chosen suitably. Then, we investigate the
instability of the sign-changeable interacting HDE model against
perturbations in BD theory. For this purpose, we study the squared
sound speed $v_s^2$ whose sign determines the stability of the
model. When $v_s^2<0$ the model is unstable against perturbation.
For future event horizon cutoff, our universe can be stable
(${v}^{2}_s>0$) depending on the model parameters. Then, we focus
on GO and Ricci cutoffs and find out that although other features
of these two cutoffs seem to be consistent with observations, they
cannot leads to stable dominated universe, except in special case
with GO cutoff. Our studies confirm that for the sign-changeable
HDE model in the setup of BD cosmology, the event horizon is the
most suitable horizon which can passes all conditions and leads to
a stable DE dominated universe.
\end{abstract}
\maketitle
%%%%%%%%%%%%%%%%%%%%%%%%%%%%%%%%%%%%%%%%%%%%%%%%%%%%%%%%%%%%%%%%%%
\section{Introduction}
The cosmological observational data from type Ia supernovae (SNIa)
\cite{Riess,Riess1,Riess2,Riess3}, the Large Scale Structure (LSS)
\cite{COL2001,COL20011,COL20012,COL20013} and the Cosmic Microwave
Background (CMB) anisotropies \cite{HAN2000,HAN20001,HAN20002},
Baryon Acoustic Oscillations (BAO) in the Sloan Sky Digital Survey
(SSDS) luminous galaxy sample \cite{Tegmark2004,Tegmark20041} and
Plank data \cite{Ade2014}, confirm that the observable universe is
nearly spatially flat, homogeneous and isotropic at large scale
and is experiencing a phase of accelerated expansion in particular
in the redshift $0.45\leq z\leq 0.9$ \cite{Cai2010}.

The provenance of this acceleration should be caused due to an
un-known energy component with negative pressure which can
overcome to gravitation of galaxy and is usually called dark
energy (DE). It is nowadays commonly accepted that DE has occupied
about $\%73$ of the total energy content of the universe and the
rest has been released to dark matter and baryonic matter. On the
other side deceleration phase is important for nucleosythesis as
well as for the structure formation. It is important to note that
we need a dynamical field in such a way that its dynamics makes at
first the deceleration phase in the early time and the
acceleration phase in the late time of the universe evolution.
This fact has motivated people for investigating dynamical DE
models. One of the dramatic candidates for dynamical models is the
HDE model which has arisen a lot of attentions
\cite{Li2004,Zhang2005,Zhang2007}. This model is based on the
holographic principle \cite{Hooft,Susskind1995} that states the
number of degrees of freedom of a system scales with its area
instead of its volume. The HDE model relates DE density to the
large length in the universe, which is usually assumed to be the
cosmic horizon. The HDE models have been investigated widely in
the literatures \cite{HDE}.

Scalar-tensor theories of gravity have been widely applied in
cosmology \cite{Fara}. The pioneering study on scalar-tensor
theories was done by Brans and Dicke (BD) several decades ago who
sought to incorporate Mach's principle into gravity
\cite{Brans1961}. In recent years, scalar tensor theories have
been reconsidered extensively, because the scalar fields appear in
different branches of theoretical physics as a consistency
condition. For example, the low energy limit of the string theory
leads to introducing a scalar degree of freedom. Since the HDE
model have a dynamical behavior, it is more reasonable to consider
it in a dynamical framework such as BD cosmology. The BD theory
also passed the observational tests in the solar system domain
\cite{Bertotti2003}. According to BD theory, the gravitational
fields are described by the metric $g_{\mu \nu}$ and a scalar
field $\varphi $ which is coupled to the gravity via a coupling
parameter $\omega $ which is restricted to a very large value
\cite{Bertotti2003,Felice2006}. The studies on the HDE model in
the framework of BD cosmology have been carried out in
\cite{Pavon2,Setare2,SHBD,other,Karami}.

On the other side, recent observations indicate that the evolution
of the two dark components of the universe is not independent and
indeed there is a mutual interaction between the Dark matter (DM)
and DE, which may solve the coincidence problem \cite{wang2016}.
However, the form of this mutual interaction can be written as $Q
=3b^2 H(\rho_{M}+\rho_{D})$, where $b^2$ is a coupling constant
and $\rho_M$ and $\rho_D$ are the energy density of DM and DE,
respectively. Clearly, the sign of this form of interaction term
cannot change during the history of the universe. While recent
investigations obliviously confirm that the sign of the
interaction term may change during the cosmic evolution, in
particular in the redshift $0.45\leq z\leq 0.9$ \cite{Cai2010}.
Wei was the first \cite{Wei2011,WEI2011} who suggested a
sign-changeable interaction term in the form $Q=q(\alpha
\dot{\rho}+3\beta H{\rho})$, where $\alpha$ and $\beta$ are
dimensionless constant and $q=-1-\dot{H}/H^2$, is the deceleration
parameter. Obviously, the interaction $Q$ can change its sign when
our universe changes from the deceleration phase $(q > 0)$ to the
acceleration $(q < 0)$. The investigation on the DE models with
sign-changeable interaction term have been carried out in
\cite{Signch}.

In the present work, we would like to investigate the HDE model
with sign-changeable interaction term in the background of BD
theory. First, we study the cosmological implications of this
model and then we explore the stability of the model against
perturbation by considering the squared sound speed
$v_s^2={dP}/{d\rho}$ whose sign determines the stability of the
model \cite{Peebles20031}. When $v_s^2<0$ the model is unstable
against perturbation. In the framework of Einstein gravity,
instability of DE models have been explored in \cite{StaDE}. While
stability of interacting HDE with GO cutoff in BD theory has been
discussed in \cite{Khodam2014}, sound instability of nonlinearly
interacting ghost dark energy have been studied in
\cite{Golchin2016}.

This paper is outlined as follows. In section \ref{GF}, we give a
brief review of the interacting HDE model in the context of BD
cosmology. In sections \ref{SG}, we study HDE in the framework of
BD theory by assuming a sign-changeable interaction term with
future horizon as system's IR cutoff. Sections \ref{GO} and
\ref{Ricci} also investigate the HDE models in BD cosmology with
GO and Ricci cutoffs, respectively. In each cases, we study the
evolution of the cosmological parameters as well as the sound
stability ${v}^{2}_{s}$ of the model. The summary of the results
is discussed in the last section.
%%%%%%%%%%%%%%%%%%%%%%%%%%%%%%%%%%%%%%%%%%%%%%%%%%%%%%%%%%%%%%%%%%%%%%%%%%
\section{Interacting HDE in BD cosmology}\label{GF}
The action of BD theory is given by
\begin{equation}
 S=\int{
d^{4}x\sqrt{g}\left(-\varphi {R}+\frac{\omega}{\varphi}g^{\mu
\nu}\partial_{\mu}\varphi \partial_{\nu}\varphi +L_M
\right)}.\label{act0}
\end{equation}
By re-defining the scalar field, $ \varphi={\phi^2 }/{8\omega} $,
we can rewrite the above action in the canonical form as
\cite{Arik2006,Arik2008}
\begin{equation}
 S=\int{
d^{4}x\sqrt{g}\left(-\frac{1}{8\omega}\phi ^2
{R}+\frac{1}{2}g^{\mu \nu}\partial_{\mu}\phi \partial_{\nu}\phi
+L_M \right)},\label{act1}
\end{equation}
where $R$ and $\phi$ are the scalar curvature and the BD scalar
field, respectively. Also, $\omega$ stands for the generic
dimensionless parameter of the BD theory and $L_M$ is the
Lagrangian of the matter. The term $\phi^2 R$, which is the
non-minimal coupling term, is replaced with the Einstein-Hilbert
term ${R}/{G}$, in such a way that $G^{-1}_{\mathrm{eff}}={2\pi
\phi^2}/{\omega}$,  where $G_{\mathrm{eff}}$ is the effective
gravitational constant. We consider a FRW universe which is
described by the line element
\begin{eqnarray}
 ds^2=dt^2-a^2(t)\left(\frac{dr^2}{1-kr^2}+r^2d\Omega^2\right). \label{metric1}
 \end{eqnarray}
where $a(t)$ is the scale factor, and $k$ is the curvature
parameter with $k = -1, 0, 1$ corresponding to open, flat, and
closed universes, respectively. Varying action (\ref{act1}) yields
the following field equations
\begin{eqnarray}
&&\frac{3}{4\omega}\phi^2\left(H^2+\frac{k}{a^2}\right)-\frac{1}{2}\dot{\phi}
^2+\frac{3}{2\omega}H
\dot{\phi}\phi=\rho_M+\rho_D,\label{FE1}\\
&&\frac{-1}{4\omega}\phi^2\left(2\frac{{\ddot{a}}}{a}+H^2+\frac{k}{a^2}\right)-\frac{1}{\omega}H
\dot{\phi}\phi -\frac{1}{2\omega}
\ddot{\phi}\phi-\frac{1}{2}\left(1+\frac{1}{\omega}\right)\dot{\phi}^2=p_{D},\label{FE2}\\
&&\ddot{\phi}+3H
\dot{\phi}-\frac{3}{2\omega}\left(\frac{{\ddot{a}}}{a}+H^2+\frac{k}{a^2}\right)\phi=0,
\label{FE3}
\end{eqnarray}
where $H=\dot{a}/a$ is the Hubble parameter, $\rho_D$ and $p_D$
are, respectively, the energy density and pressure of dark energy.
There are not compelling  reason for choice of $\phi $, thus we
assume the BD field as $\phi=\phi_0 a^{\alpha}(t)$, which leads to
the following relations
\begin{equation}
\frac{\dot{\phi }}{\phi }=\alpha H,\   \   \   \frac{\ddot{ \phi
}}{\phi } =\alpha^{2}H^{2}+\alpha\dot{H},\   \   \
\frac{\ddot{\phi }}{\dot{\phi }}=\left(\alpha+\frac{
\dot{H}}{H^{2}}\right)H. \label{square}
\end{equation}
We further assume the energy density of the HDE can be written as
\begin{eqnarray}\label{rhoD}
\rho_{D}=\frac{3 c^2\phi^2 }{4\omega L^2},
\end{eqnarray}
where $\phi^2={\omega}/{2\pi G_{\mathrm{eff}} }$. In the limiting
case where $G_{\mathrm{eff}}$ reduces to $G$, the energy density
(\ref{rhoD}) reduces to the energy density of HDE in standard
cosmology,
\begin{eqnarray}\label{rhoD1}
\rho_{D}=\frac{3 c^2 }{8\pi  G L^2}=\frac{3 c^2 m_p ^2 }{ L^2}.
\end{eqnarray}
If we define the critical energy density as
\begin{eqnarray}\label{rhocr}
\rho_{\mathrm{cr}}=\frac{3\phi^2 H^2}{4\omega},
\end{eqnarray}
then the dimensionless density parameters can be written
\begin{eqnarray}
\Omega_M&=&\frac{\rho_M}{\rho_{\mathrm{cr}}}=\frac{4\omega\rho_M}{3\phi^2
H^2}, \label{Omegam} \\
\Omega_k&=&\frac{\rho_k}{\rho_{\mathrm{cr}}}=\frac{k}{H^2 a^2},\label{Omegak} \\
\Omega_D&=&\frac{\rho_D}{\rho_{\mathrm{cr}}}=\frac{4\omega\rho_D}{3\phi^2
H^2}. \label{OmegaD}
\end{eqnarray}
For the  FRW universe filled with DE and DM, with mutual
interaction, the semi-conservation equations are as follow
\begin{eqnarray}
&&
\dot{\rho}_D+3H\rho_D(1+w_D)=-Q,\label{cons1}\\
&&\dot{\rho}_{M}+3H\rho_{M}=Q, \label{cons2}
\end{eqnarray}
where $Q$ is the interaction term which we assume has the form $Q
=3b^2 q H(\rho_{M}+\rho_{D})$ \cite{Wei2011,chimen1,chimen2},
where $ b^2$ is a coupling constant and $q$ is the deceleration
parameter,
\begin{equation}\label{deceleration}
q=-\frac{\ddot{a}}{a H^2}=-1-\frac{\dot{H}}{H^2}.
\end{equation}
Combining Eqs. (\ref{square}) and (\ref{rhocr}) with Friedmann Eq.
(\ref{FE1}), we arrive at
\begin{equation}\label{rhocr1}
\rho_{\rm cr}+\rho_{k}=\rho_{M}+\rho_{D}+\rho_{\phi},
\end{equation}
where we have defined
\begin{equation}\label{rhocr2}
\rho_{\phi}\equiv\frac{1}{2} \alpha H^2 \phi ^2
\left(\alpha-\frac{3}{\omega}\right).
\end{equation}
Dividing Eq.(\ref{rhocr1}) by $ \rho_{\rm cr}$, this equation can
be rewritten as
\begin{equation}\label{OmegaD1}
\Omega_{m}+\Omega_{D}+\Omega_{\phi}=1+\Omega_{k},
\end{equation}
where
\begin{equation}\label{Omegaphi}
\Omega_{\phi}=\frac{\rho_{\phi}}{\rho_{\rm cr}}=-2\alpha
\left(1-\frac{\alpha\omega}{3}\right).
\end{equation}
Next, we introduce the ratio of the energy densities, $r$, which
can be written
\begin{equation}\label{r}
r=\frac{\Omega_{m}}{\Omega_{D}}=-1+\frac{1}{\Omega_{D}}\left[1+\Omega_{k}+2\alpha
\left(1-\frac{\alpha \omega}{3}\right)\right].
\end{equation}
The idea for investigating the stability of any DE model comes
from the perturbation theory. For this purpose, we assume a small
perturbation in the background energy density. We are interested
in checking whether the perturbation grows with time or it will
collapse. In the linear perturbation theory, the perturbed energy
density of the background can be written as
\begin{equation}\label{pert1}
\rho(t,x)=\rho(t)+\delta\rho(t,x),
\end{equation}
where $\rho(t)$ is unperturbed background energy density. The
energy conservation equation ($\nabla_{\mu}T^{\mu\nu}=0$) yields
\cite{Peebles20031}
\begin{equation}\label{pert2}
\delta\ddot{\rho}=v_s^2\nabla^2\delta\rho(t,x),
\end{equation}
where $v_s^2={dP}/{d\rho}$ is the squared of the sound speed.
There are two kind of solutions for Eq. (\ref{pert2}). In the
first case  where $v_s^2>0$,  Eq. (\ref{pert2}) becomes an
ordinary wave equation which have a wave solution in the form
$\delta \rho=\delta\rho_0e^{-i\omega t+i\vec{k}.\vec{x}}$.
Clearly, in this case the density perturbations propagates with
time and the system is stable. In the second case where $v_s^2<0$,
the frequency of the oscillations becomes pure imaginary and the
density perturbations will grow with time as $\delta
\rho=\delta\rho_0e^{\omega t+i\vec{k}.\vec{x}}$. This implies a
possible emergency of instabilities in the background. Therefore,
the sign of ${v}^{2}_{s}$ plays a crucial role in determining the
stability of DE model. If ${v}^{2}_{s}<0$ (${v}^{2}_{s}>0$) it
means that we have the classical instability (stability)of a given
perturbation. The quantity ${v}^{2}_{s}$ for the FRW universe is
given by
\begin{equation}\label{stable}
{v}^{2}_{s}=\frac{\dot P}{\dot\rho} =\frac{\dot\rho_{D}
w_{D}+\rho_{D}\dot w_D}{\dot\rho_{D}(1+r)+\rho_{D}\dot r},
\end{equation}
where $P=P_D$ is the pressure of DE and $\rho=\rho_{M}+\rho_{D}$
is the total energy density of DE and DM.
%%%%%%%%%%%%%%%%%%%%%%%%%%%%%%%%%%%%%%%%%%%%%%%%%%%%%%%%%%%%%%%%%%%%%%%
\section{Sign-Changeable HDE in BD theory with future horizon cutoff}\label{SG}
At the beginning, we consider the future event horizon as
system\textquoteright s IR cutoff, which is defined as
\begin{eqnarray}\label{cutoffEvent}
L=R_{h}=a(t)\int_{t}^{\infty}{\frac{dt}{a(t)}},
\end{eqnarray}
from which we can get
\begin{eqnarray}\label{rhoDh}
\dot{R_{h}}=H R_{h}-1,
\end{eqnarray}
and hence from Eqs. (\ref{rhoD}) and (\ref{OmegaD}) we can obtain
\begin{eqnarray}\label{rhoDEvent}
\rho_{D}=\frac{3 c^2\phi^2 }{4\omega R_{h}^2},
\end{eqnarray}
\begin{eqnarray}\label{OmegaDEvent}
\Omega_D=\frac{\rho_D}{\rho_{\mathrm{cr}}}=\frac{c^2}{ H^2
R_{h}^2}.
\end{eqnarray}
Taking the time derivative of the energy density $\rho_{D}$ in Eq.
(\ref{rhoDEvent}) and simultaneously using Eqs. (\ref{square}),
(\ref{rhoDh}) and (\ref{OmegaDEvent}) we arrive at
\begin{equation}\label{dotrho}
\dot\rho_{D}=2H\rho_{D}\left(\alpha-1+\frac{\sqrt{\Omega_{D}}}{c}\right).
\end{equation}
\begin{figure}[htp]
\begin{center}
\includegraphics[width=8cm]{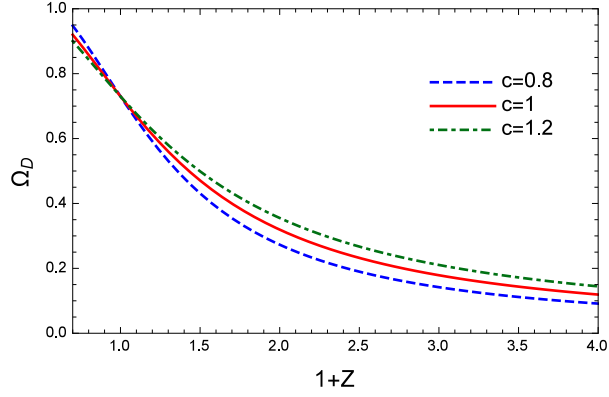}
\caption{Evolution of $\Omega_D$ versus redshift parameter $z$ for
the sign-changeable interacting HDE with future event horizon as
IR cutoff in BD cosmology. Here, we have taken
$\Omega^{0}_D=0.73$, $\alpha=10^{-4}$, $\omega=10^4$ and
$b^2=0.1$. }\label{Omega-z1}
\end{center}
\end{figure}

\begin{figure}[htp]
\begin{center}
\includegraphics[width=8cm]{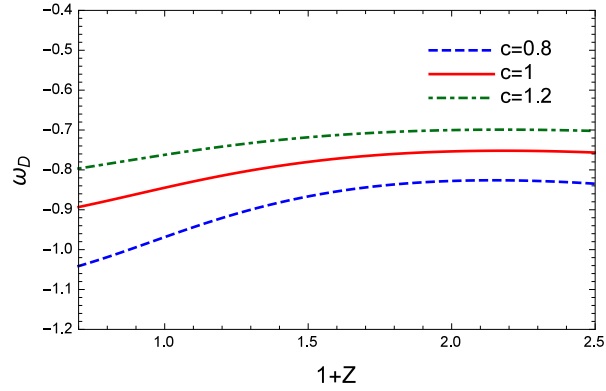}
\caption{Evolution of $w_D$ versus redshift parameter $z$ for the
sign-changeable interacting HDE with future event horizon as IR
cutoff in BD cosmology. Here, we have taken $\alpha=10^{-4}$,
$\omega=10^4$ and $b^2=.1$ as the initial condition.
}\label{EoS-z1}
\end{center}
\end{figure}

\begin{figure}[htp]
\begin{center}
\includegraphics[width=8cm]{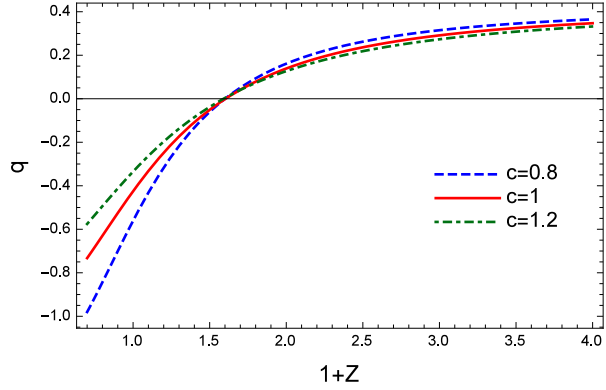}
\caption{Evolution of the deceleration parameter $q$ against
redshift parameter $z$ for the sign-changeable interacting HDE
with future event horizon as IR cutoff in BD cosmology. Here, we
have taken $\alpha=10^{-4}$, $\omega=10^4$ and $b^2=.1$ as the
initial condition. }\label{q-z1}
\end{center}
\end{figure}
Next, we determine the EoS parameter with inserting Eq.
(\ref{dotrho}) in semi-conservation law (\ref{cons1}). We find
\begin{equation}\label{omega1}
w_{D}=-1-b^2 q (1+r)-\frac{2}{3}
\left(\alpha-1+\frac{\sqrt{\Omega_{D}}}{c}\right).
\end{equation}
The deceleration parameter $ q $ can be obtained by dividing
Eq.(\ref{FE2}) by $H^2$ and using Eqs. (\ref{square}),
(\ref{deceleration}) and (\ref{OmegaDEvent}),
\begin{equation}\label{deceleration1}
q=\frac{1}{2\alpha+2}\left[2\alpha^2(\omega+2)+2\alpha+1+3
\Omega_{D} w_{D}\right].
\end{equation}
Substituting $\omega_{D}$ from (\ref{omega1}) in above relation,
we reach
\begin{equation}\label{deceleration2}
q=\frac{c(1+2\alpha)\gamma_0 -c(2\alpha c+2\sqrt{\Omega_{D}})\Omega_{D}}{2c\gamma_0+3b^2 c(1+r)\Omega_{D}},
\end{equation}
where $\gamma_0=1-{2 \omega \alpha^{2}}/{3}-2\alpha$. Taking the
time derivative of Eq. (\ref{OmegaDEvent}) and using Eq.
(\ref{rhoDh}) as well as the fact that
$\dot\Omega_{D}={\Omega}^{\prime}_{D} H$, we can obtain the
equation of motion for $\Omega_{D}$ as
\begin{equation}\label{OmegaD3}
{\Omega}^{\prime}_{D}=2\Omega_{D}\left(q+\frac{\sqrt{\Omega_{D}}}{c}\right),
\end{equation}
where the dot and the prime indicate differentiation with respect
to the cosmic time and  $x=\ln{a}$, respectively. Substituting $q$
from relation (\ref{deceleration2}) in Eq. (\ref{OmegaD3}), we can
Setting $n=0$ $(\omega\rightarrow\infty)$, which is the limiting
case of Einstein gravity \cite{Lu2012}, the obtained results in
Eqs. (\ref{omega1}), (\ref{deceleration2}) and (\ref{OmegaD3})
reduce to their respective expressions in Einstein gravity
\cite{Abdollahi2017}.
\begin{figure}[htp]
\begin{center}
\includegraphics[width=8cm]{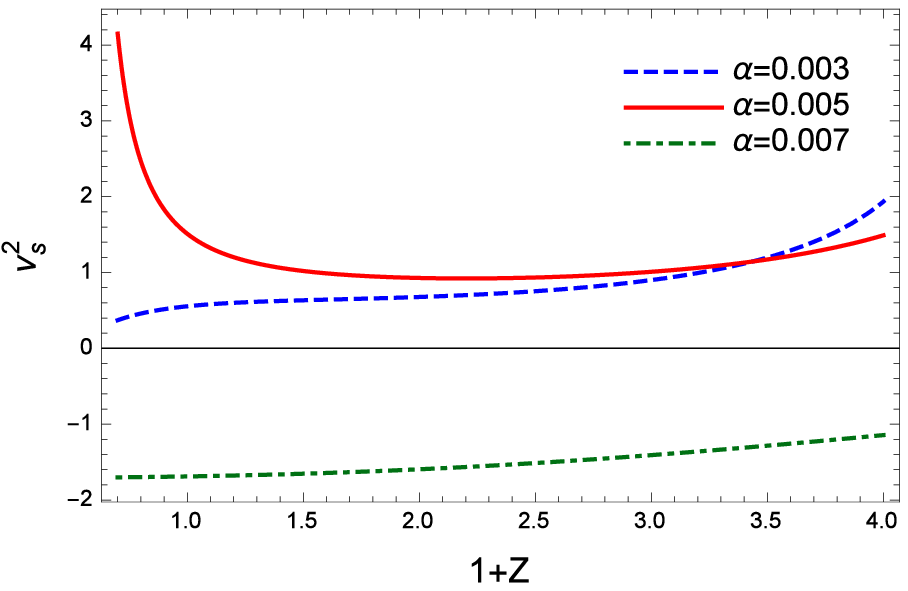}
\includegraphics[width=8cm]{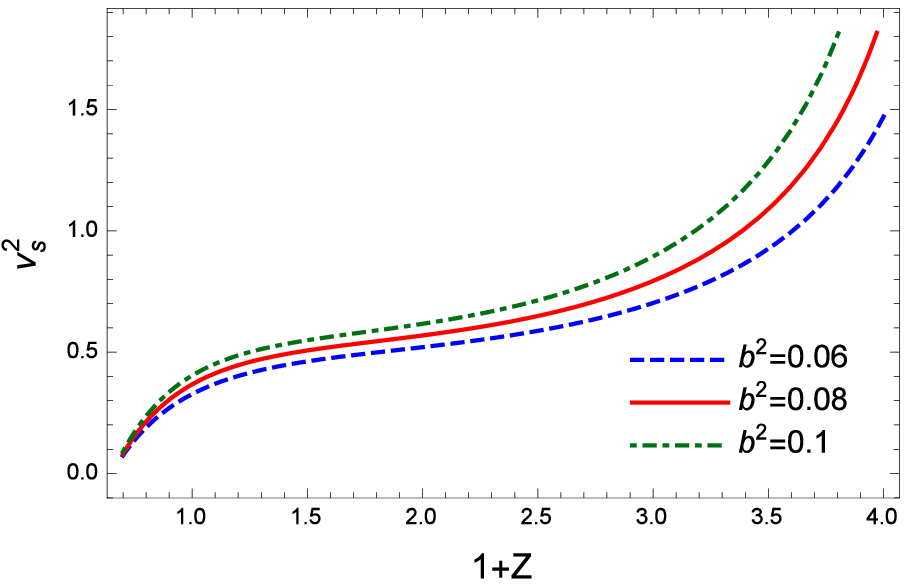}
\caption{Evolution of the squared of sound speed ${v}^{2}_{s}$
against redshift parameter $z$ for the sign-changeable interacting
HDE with Future cutoff  in BD cosmology. Here, we have taken
$c^2=1$, $b^2=0.1$ and $\omega=10^4$ in the left panel and
$\alpha=10^{-4}$, $\omega=10^4$ and $c^2=1$  in the right panel,
as the initial condition, respectively}\label{S-z1}
\end{center}
\end{figure}

\begin{figure}[htp]
\begin{center}
\includegraphics[width=8cm]{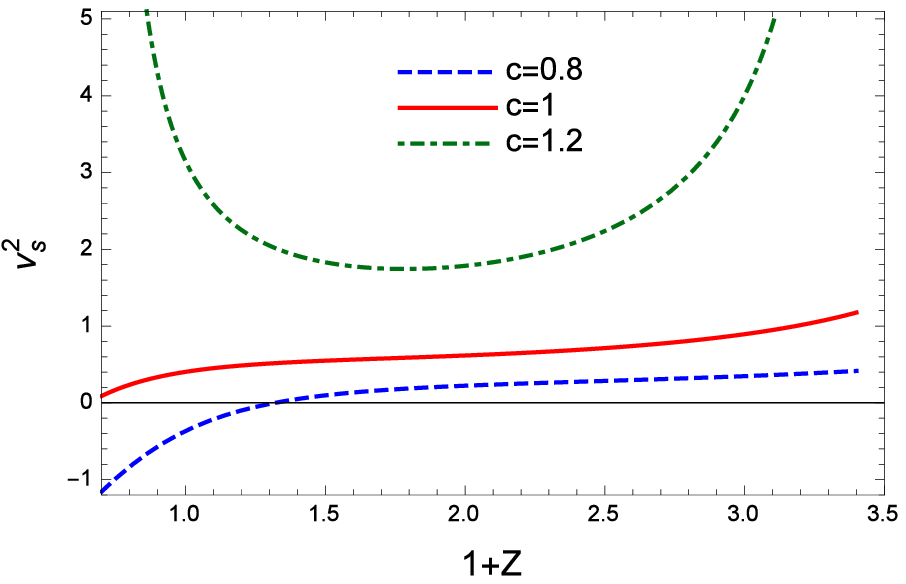}
\includegraphics[width=8cm]{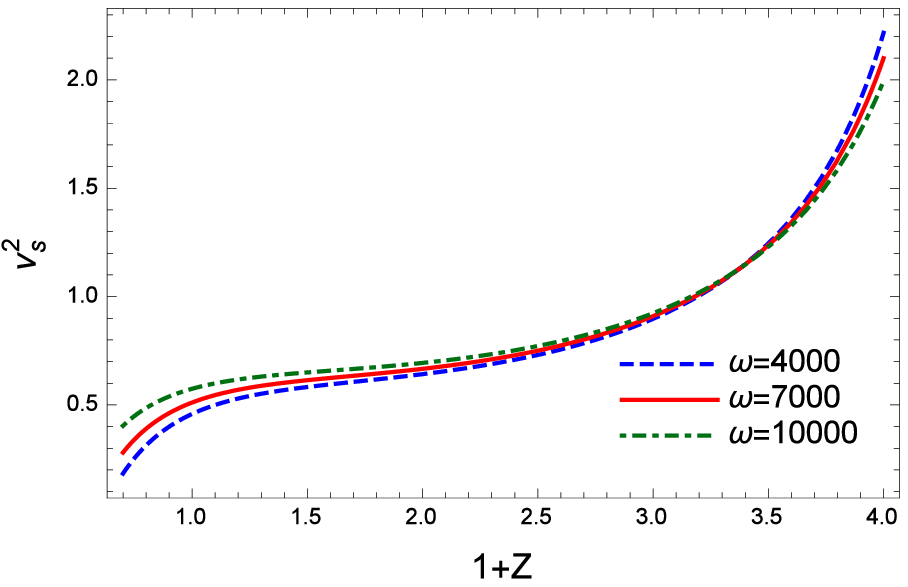}
\caption{Evolution of the squared of sound speed ${v}^{2}_{s}$
against redshift parameter $z$  for the sign-changeable
interacting HDE with Future cutoff  in BD cosmology. Here, we have
taken $\alpha=10^{-4}$, $\omega=10^4$ and $b^2=0.1$ in the left
panel and  $\alpha=.003$, $b^2=0.1$ and $c^2=1$  in the right
panel, as the initial condition, respectively}\label{S-z2}
\end{center}
\end{figure}
Substituting Eqs. (\ref{r}), (\ref{dotrho}) and (\ref{omega1}) in
Eq. (\ref{stable}), we can obtain the explicit expression for
${v}^{2}_{s}$. Since this expression is too long, we shall not
present it here, instead we study the evolution of ${v}^{2}_{s}$
via figures. To illustrate the cosmological consequences of the
HDE with sign-changeable interaction term in BD cosmology, we also
plot the density parameter $\Omega_{D}$, and the EoS parameter
$w_{D}$ and the deceleration parameter $q$ and the squared sound
speed ${v}^{2}_{s}$ in Figs.~\ref{Omega-z1}- \ref{S-z2}. In
Fig.~\ref{Omega-z1}, as we expect, it is seen that in early time
of universe ($1+z\rightarrow\infty$) we have $\Omega_D\rightarrow
0$, while at the late time where ($1+z\rightarrow 0$), we have
$\Omega_D\rightarrow 1$. From Fig.~\ref{EoS-z1} one can clearly
see that for $ c\lesssim 1$ the EoS parameter $w_D$ can cross the
phantom line and when $c\geq 1$, we always have $w_D >-1$. As it
is obvious from Fig.~\ref{q-z1}, for all values of $c$, the
deceleration parameter $q$ transits from deceleration ($q>0$) in
the early time to acceleration ($q< 0$) in the last time around
$z\approx 0.6$. The evolution of ${v}^{2}_{s}$ versus $z$ for the
different parameters $\alpha$, $b^2$, $c$ and $\omega$ are plotted
in Figs.~\ref{S-z1} and \ref{S-z2}. Graphical analysis of
${v}^{2}_{s}$ shows that in Fig.~\ref{S-z1} (left panel) the
sign-changeable HDE in BD theory could be stable for suitable
values of $\alpha$. Also, in Fig.~\ref{S-z2} by increasing $c$,
the squared sound speed, ${v}^{2}_{s}$, is positive which implies
that the sign-changeable interacting HDE with the future cutoff in
BD cosmology can be stable. Since at the present time, our
Universe is experiencing a phase of accelerated expansion, thus we
need to find a model which respects the stability condition around
the present time. Obviouslyæ our present model passes all
conditions.
%%%%%%%%%%%%%%%%%%%%%%%%%%%%%%%%%%%%%%%%%%%%%%%%%%%%%%%%%%%%%%%%%%%%%%%%%%%%%%%%
\section{Sign-Changeable HDE with GO Cutoff in BD theory}\label{GO}
The energy density of HDE in BD theory with GO cutoff
\cite{Granda2008}, is given by
\begin{eqnarray}\label{rhoGo}
\rho_{D}=\frac{3 \phi^2 }{4\omega }(\gamma H^2+\beta \dot{H}).
\end{eqnarray}
Dividing Eq. (\ref{FE1}) by $H^2$  and using Eq.(\ref{square}) as
well as the above relation, we obtain
\begin{eqnarray}\label{dotH}
\frac{\dot{H}}{H^2}=\frac{1-{2\omega
\alpha^{2}}/{3}+2\alpha}{\beta (1+r)}-\frac{\gamma}{\beta}.
\end{eqnarray}
From Eq.(\ref{deceleration}), it follows that
\begin{eqnarray}\label{decelerationGo}
q=-1+\frac{\gamma}{\beta}-\frac{1-{2\omega
\alpha^{2}}/{3}+2\alpha}{\beta (1+r)}.
\end{eqnarray}
The EoS parameter $w_{D}$ of the HDE in BD theory is given
\begin{equation}\label{EoSGo}
w_{D}=\frac{1}{3\left[\gamma
-\beta(1+q)\right]}\left[(2q-1)-4\alpha-4
\alpha^{2}+2\alpha(1+q)-2\alpha^{2}\omega \right].
\end{equation}
Dividing Eq. (\ref{FE2}) by $ H^2$ and using of relation
(\ref{rhoGo}), after inserting Eq. (\ref{decelerationGo}) in
Eq.(\ref{EoSGo}), we arrive at
\begin{eqnarray}\label{EoSGo1}
w_{D}=\frac{6+(9\beta-6\gamma)(1+r)+6\alpha
\left[3+2\beta(1+r)-\gamma(1+r)\right]-4\alpha^{2}\omega+2\alpha^{2}
\left[6-2\omega+3\beta
(1+r)(2+\omega)\right]}{3\beta[-3+2\alpha(-3+\alpha\omega)]}.\nonumber\\
\end{eqnarray}
Using relation $\Omega_D={\rho_D}/{\rho_{\mathrm{cr}}}$, and Eq.
(\ref{rhoGo}), we get
\begin{eqnarray}\label{OmegaGo}
\Omega_D=\gamma+\beta\frac{\dot{H}}{H^2}.
\end{eqnarray}
Taking the derivative of this relation respect to the cosmic time
$t$ and using Eq. (\ref{dotH}), we find
\begin{eqnarray}\label{OmegadotGo}
\dot\Omega_D=-\frac{\dot{r}\left(1-\frac{2\omega
n^2}{3}+2n\right)}{ (1+r)^2},
\end{eqnarray}
where $\dot\Omega_D=H{\Omega}^{\prime}_{D}$, and $\dot r$ can be
obtained using $r={\rho_m}/{\rho_D}$ as well as the conservation
equations,
\begin{eqnarray}\label{rdotGo}
\dot r=3H\left[b^2 q(1+r)^2+\omega_{D}r\right].
\end{eqnarray}
When $\alpha=0$, Eqs. (\ref{decelerationGo}), (\ref{EoSGo1}) and
(\ref{OmegadotGo}) reduce to their respective expressions in flat
standard cosmology \cite{Abdollahi2017}. Computing $\dot w_{D}$
and using Eq.(\ref{rdotGo}), after replacing in relation
(\ref{stable}), we can investigate the squared speed of sound $
{v}^{2}_{s}$. Again, for the economic reason, we do not bring the
explicit expression of ${v}^{2}_{s}$, instead we focus on its
behaviour via figures.

\begin{figure}[htp]
\begin{center}
\includegraphics[width=8cm]{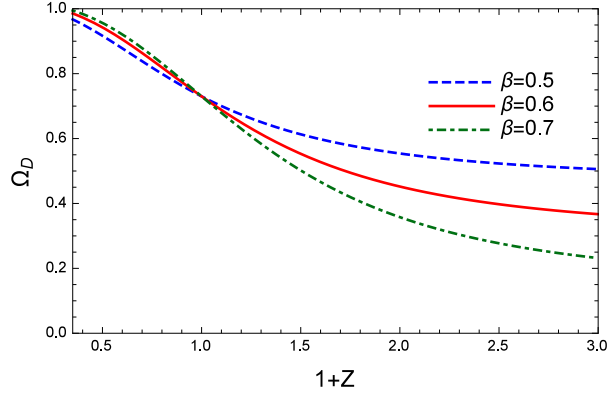}
\caption{ Evolution of $\Omega_D$ versus $1+z$ parameter for the
sign-changeable interacting HDE with GO cutoff  in BD cosmology.
Here, we have taken $\Omega_D(z=0)=0.73$,$\alpha=.003$,
$\omega=10^4$, $b^2=0.1$ and $\gamma=1.2$ as the initial
conditions.}\label{OmegaG-z1}
\end{center}
\end{figure}

\begin{figure}[htp]
\begin{center}
\includegraphics[width=8cm]{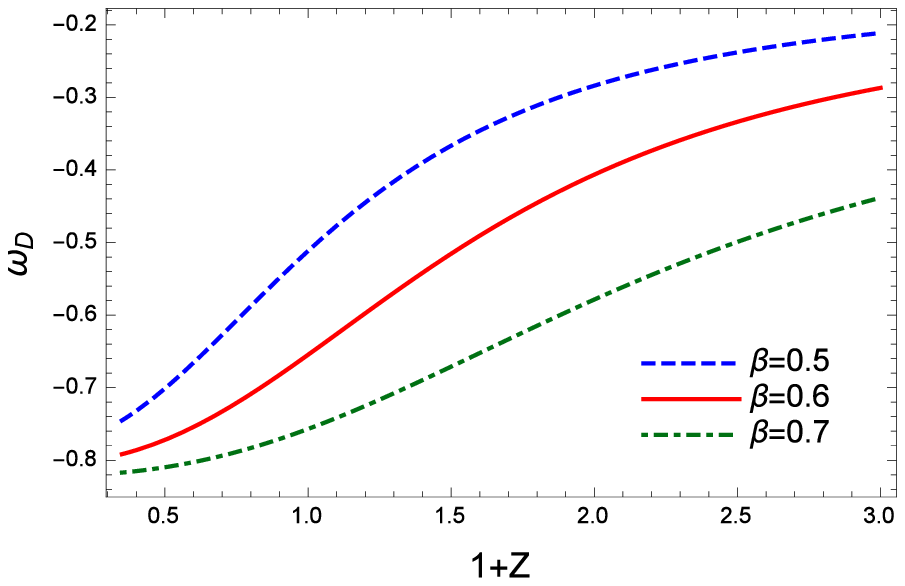}
\includegraphics[width=8cm]{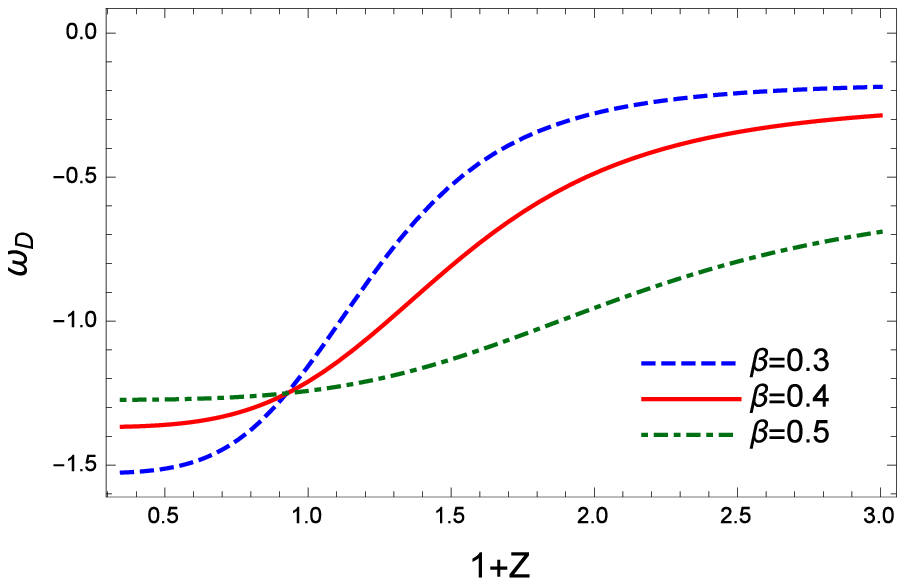}
\caption{Evolution of $w_D$ versus  $z$ for HDE with GO cutoff. In
the left panel  we have taken $\alpha=10^{-4}$, $\omega=10^4$,
$b^2=.1$ and $\gamma=1.2$ and in the right panel $\alpha=10^{-4}$,
$\omega=10^4$, $b^2=0.1$ and $\gamma=0.8$ as the initial
condition, respectively. }\label{EoSG-z1}
\end{center}
\end{figure}
\begin{figure}[htp]
\begin{center}
\includegraphics[width=8cm]{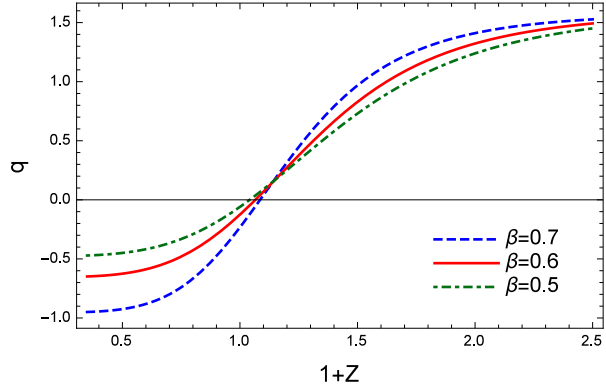}
\caption{Evolution of the  deceleration parameter $q$ against $z$
for HDE with GO cutoff in the BD cosmology. We have taken
$\alpha=10^{-4}$, $\omega=10^4$, $b^2=0.1$ and $\gamma=1.2$ as the
initial condition.}\label{qG-z1}
\end{center}
\end{figure}

\begin{figure}[htp]
\begin{center}
\includegraphics[width=8cm]{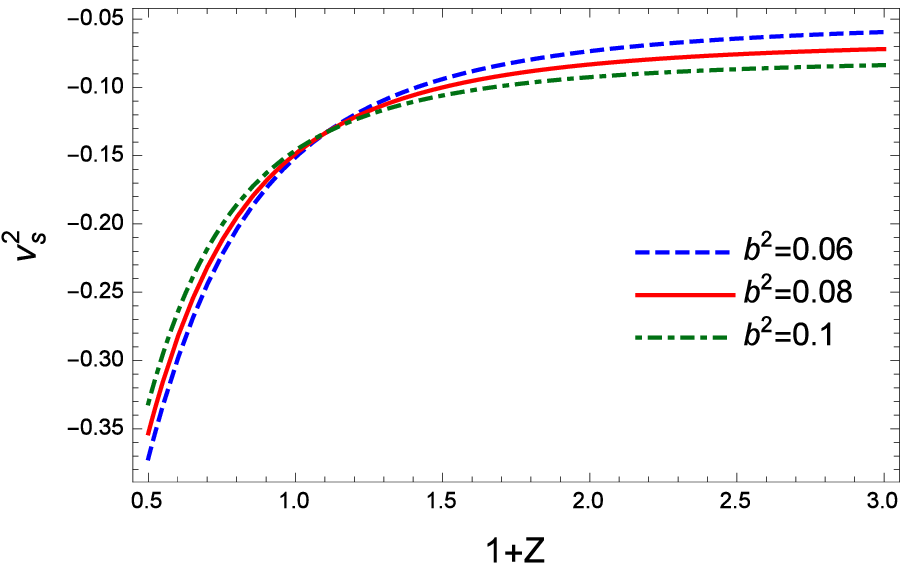}
\includegraphics[width=8cm]{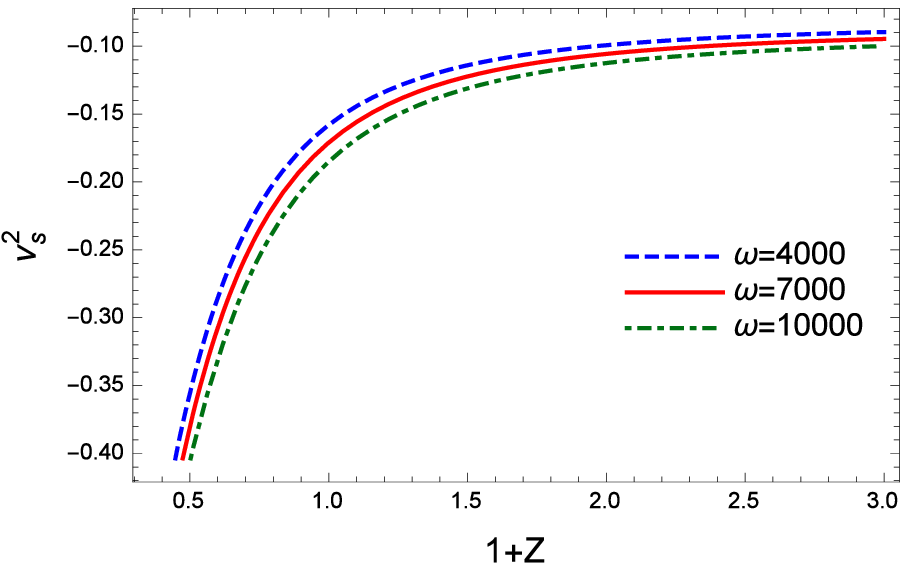}
\caption{Evolution of the squared of sound speed ${v}^{2}_{s}$
against $z$ for the sign-changeable interacting HDE with GO
cutoff  in BD cosmology. When  $\alpha=10^{-4}$, $\gamma=1.2$,
$\beta=.5$ and $\omega=10^4$ in the left panel and $\alpha=.003$,
$b^2=0.1$,  $\gamma=1.2$ and $\beta=0.5$  in the right panel, as
the initial condition, respectively.}\label{SG-z1}
\end{center}
\end{figure}

\begin{figure}[htp]
\begin{center}
\includegraphics[width=8cm]{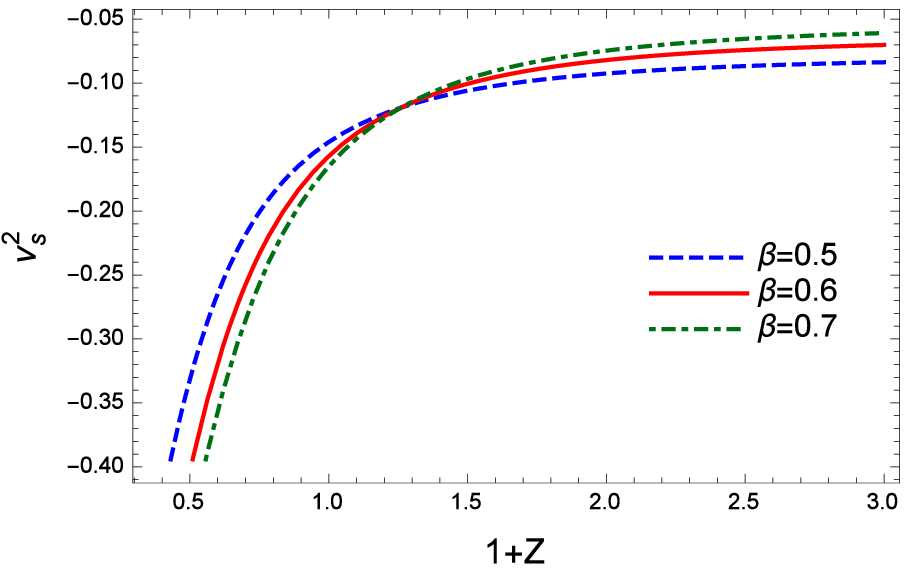}
\includegraphics[width=8cm]{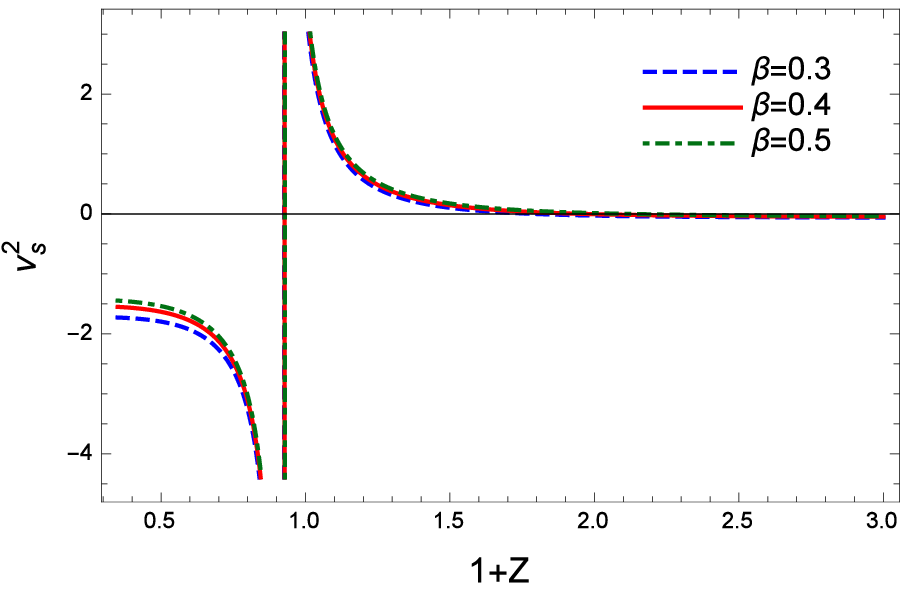}
\caption{Evolution of the squared of sound speed ${v}^{2}_{s}$
versus $z$ for the sign-changeable interacting HDE with GO cutoff
in BD cosmology. When $b^2=0.1$, $\alpha=10^{-4}$, $\gamma=1.2$,
and $\omega=10^4$ in the left panel and  $\alpha=10^{-4}$,
$\omega=10^4$ $b^2=0.1$ and $\gamma=0.8$  in the right panel, as
the initial condition, respectively.}\label{SG-z2}
\end{center}
\end{figure}
The behavior of $\Omega_{D}$ against redshift parameter for HDE
with GO cutoff and in the setup of BD theory has been plotted in
Fig.~\ref{OmegaG-z1}. We find that at the late time where the DE
dominates we have $\Omega_D\rightarrow 1$, while
$\Omega_D\rightarrow 0$ at the early time. The graphical behavior
of the EoS parameter, which is given in Eq.(\ref{EoSGo}), also
plotted in Fig.~\ref{EoSG-z1}, showing that for $\gamma=0.8$ the
EoS parameter can cross the phantom line. The behavior of the
deceleration parameter $q$ has also plotted in Fig.~\ref{qG-z1}
which indicates that our Universe has a phase transition from
deceleration to an acceleration. The behaviour of ${v}^{2}_{s}$ is
plotted against $z$ in Fig.~\ref{SG-z1}, (left panel) for
different values of the coupling parameter $b^2$ and (right panel)
for different values of $\omega$. From these figure we see that
increasing $b^2$, leads to more instability against perturbations.
Also, the squared sound speed is studied in Fig.~\ref{SG-z2} for
different values of $\beta$  by assuming $\gamma=1.2$ (left panel)
and $\gamma=0.8$ (right panel), which reveals that for $\gamma>1$
this model is instable, whereas we can obtain an stable universe
by taking $\gamma< 1$ ($\gamma=0.8$).
%%%%%%%%%%%%%%%%%%%%%%%%%%%%%%%%%%%%%%%%%%%%%%%%%%%%%%%%%%%%%%%%%%%%%%%%%%%%%%%%%%%%%%%%
\section{Sign-Changeable HDE in BD theory with Ricci Cutoff}\label{Ricci}
In this section we choose the Ricci scalar $R$ as IR cutoff
\cite{Gao}, which is given for the flat FRW Universe as
\begin{eqnarray}\label{Ricci1}
R=6(\dot{H}+2H^2).
\end{eqnarray}
Thus, the HDE density is written as
\begin{eqnarray}\label{rhoRicci}
\rho_{D}=\frac{3 c^2 \phi^2 }{4\omega }\left(\dot{H}+2H^2\right).
\end{eqnarray}
Following the method of the previous section we can find
deceleration parameter $q$ by dividing Eq.(\ref{FE1}) by $H^2$ and
using relation (\ref{rhoRicci}). We find
\begin{eqnarray}\label{dotHRicci}
\frac{\dot{H}}{H^2}=-2+\frac{1-{2\omega
\alpha^{2}}/{3}+2\alpha}{c^2(1+r)}.
\end{eqnarray}
\begin{figure}[htp]
\begin{center}
\includegraphics[width=8cm]{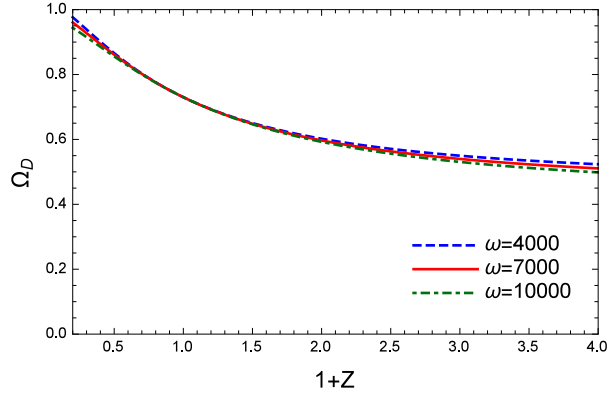}
\caption{Evolution of $\Omega_D$ versus redshift parameter $z$ for
HDE with Ricci cutoff in the BD cosmology when
$\Omega_D(z=0)=.73$,$\alpha=.003$, $c^2=.8$ and
$b^2=.1$.}\label{OmegaR-z1}
\end{center}
\end{figure}
Substituting Eq.(\ref{dotHRicci}) in relation
(\ref{deceleration}), we have
\begin{eqnarray}\label{qRicci}
q=1-\frac{1-{2\omega \alpha^{2}}/{3}+2\alpha}{c^2(1+r)}.
\end{eqnarray}
\begin{figure}[htp]
\begin{center}
\includegraphics[width=8cm]{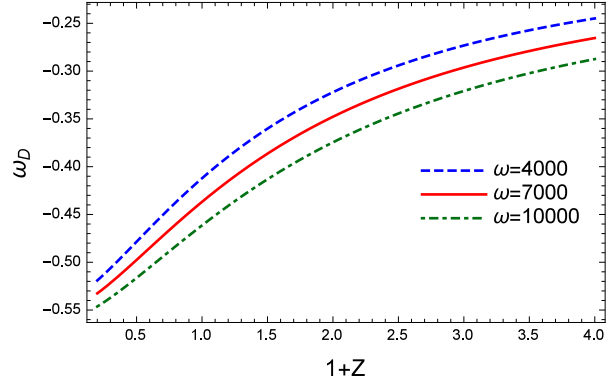}
\caption{The evolution of $w_D$ versus redshift parameter $z$ for
HDE with Ricci cutoff in the BD cosmology when $\alpha=.003$,
$c^2=.8$ and $b^2=.1$.}\label{EoSR-z1}
\end{center}
\end{figure}

\begin{figure}[htp]
\begin{center}
\includegraphics[width=8cm]{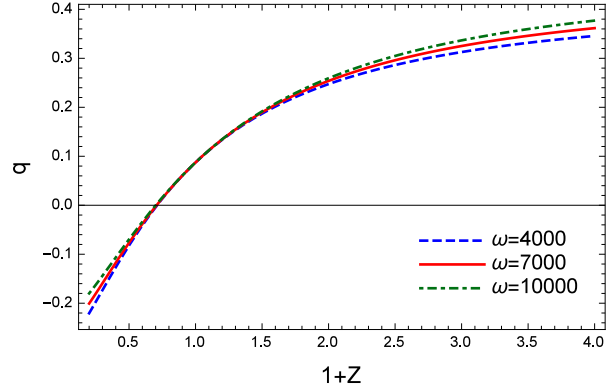}
\caption{The evolution of the  deceleration parameter $q$ against
redshift parameter $z$ for HDE with Ricci cutoff in the BD
cosmology when $\alpha=.003$, $c^2=.8$ and
$b^2=0.1$.}\label{qR-z1}
\end{center}
\end{figure}
The EoS parameter $w_ {D}$ can be obtained by dividing Eq.
(\ref{FE2}) by $ H^2$ and using relation (\ref{rhoRicci}). We find
\begin{eqnarray}\label{EosRicci}
w_{D}=-\frac{-2q+1+4\alpha+2\left[\alpha^{2}-\alpha(1+q)\right]+2\alpha^{2}(1+\omega)}{3c^2(1-q)}.
\end{eqnarray}
Substituting $q$ from (\ref{qRicci}) in Eq. (\ref{EosRicci}) we
reach
\begin{eqnarray}\label{EosRicci1}
w_{D}=\frac{3c^2(1+r)[-1+2\alpha^{2}
(2+\omega)-(1+2\alpha)]-2(1+\alpha)[2\alpha(-3+\alpha\omega)-3]}{3
c^2 [2\alpha(-3+\alpha \omega)-3]}.
\end{eqnarray}
Taking the time derivative of relation
$\Omega_D={\rho_D}/{\rho_{\mathrm{cr}}}$ with respect to the
cosmic time $t$ and using Eqs.(\ref{rhoRicci}) and
(\ref{dotHRicci}), we can find
\begin{eqnarray}\label{OmegadotRicci}
\dot\Omega_D=-\frac{\dot{r}(1-{2\omega \alpha^{2}}/{3}+2\alpha)}{
(1+r)^2}.
\end{eqnarray}
In what follow, we shall study the squared speed of sound $
{v}^{2}_{s}$ by computing $\dot w_{D}$ and replacing in relation
(\ref{stable}).

\begin{figure}[htp]
\begin{center}
\includegraphics[width=8cm]{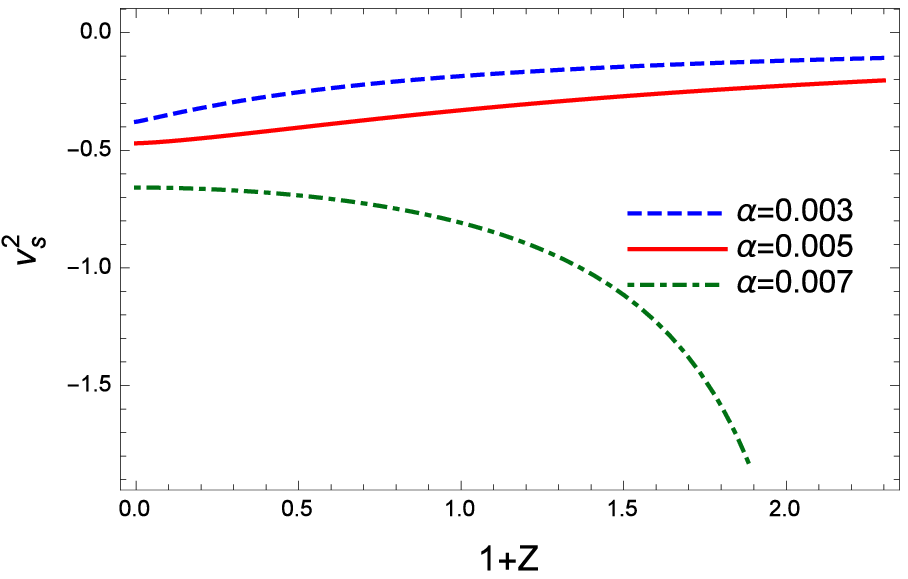}
\includegraphics[width=8cm]{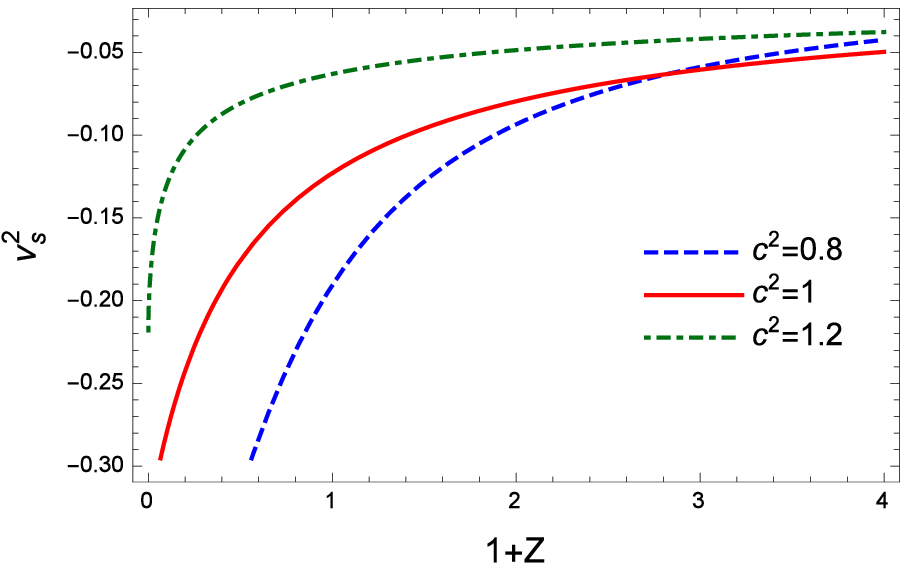}
\caption{Evolution of the squared of sound speed ${\nu}^{2}_{s}$
versus redshift parameter $z$ for the sign-changeable interacting
HDE with Ricci cutoff  in BD cosmology. When $b^2=.01$, $c^2=1$
and $\omega=10^4$ in the left panel and  $\alpha=10^{-4}$,
$\omega=10^4$ and $b^2=.01$ in the right panel, as the initial
condition, respectively.}\label{SR-z1}
\end{center}
\end{figure}

\begin{figure}[htp]
\begin{center}
\includegraphics[width=8cm]{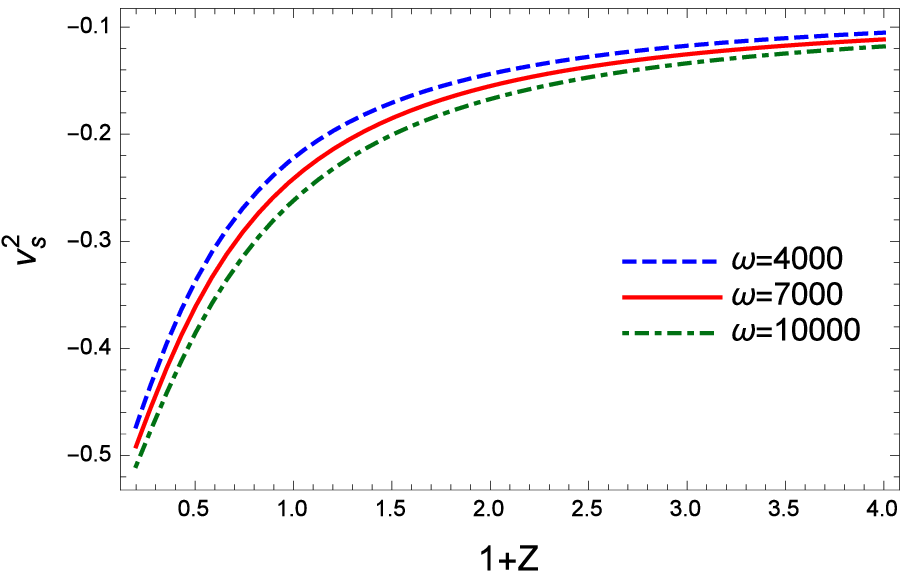}
\includegraphics[width=8cm]{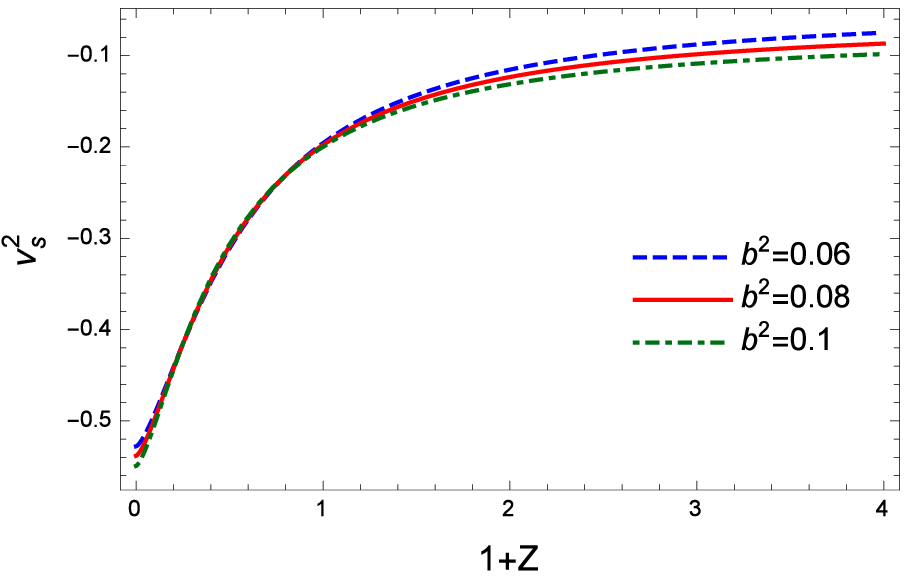}
\caption{Evolution of the squared of sound speed ${\nu}^{2}_{s}$
versus redshift parameter $z$ for the sign-changeable interacting
HDE with Ricci cutoff  in BD cosmology. Here, we have taken
$b^2=.1$, $c^2=.8$ and $\alpha=.003$ in the left panel and
$\alpha=10^{-4}$, $\omega=10^4$ and $c^2=.8$ in the right panel,
as the initial condition, respectively.}\label{SR-z2}
\end{center}
\end{figure}
Choosing the same set of parameters, we start our analysis by
plotting the behavior of all cosmological parameters such as
$\Omega_{D}$, $w_{D}$ and $q$ in Figs.~\ref{OmegaR-z1},
\ref{EoSR-z1} and \ref{qR-z1}. The main results of this figures
are as follows. ($i$) At late time, the EoS parameter cannot cross
the phantom line and we have always $w_{D}>-1$, $q$ is in
acceleration phase and $\Omega_D\rightarrow 1$. ($ii$) At early
time $w_{D}$ increases with increasing the redshift parameter, $q$
is in deceleration phase and $\Omega_D\rightarrow 0$. Having the
squared sound speed at hand, we can discuss the stability of this
model against perturbations. In summary, for this model
($\Omega_{D}$, $w_{D}$ and $q$) seem to be consistent with
observations. Besides, Figs.~\ref{SR-z1} and \ref{SR-z2} indicate
that ${v}^{2}_{s}$ remains negative which shows a sign of
instability.
%%%%%%%%%%%%%%%%%%%%%%%%%%%%%%%%%%%%%%%%%%%%%%%%%%%%%%%%%%%%%%%%%%%%%
\section{Closing remarks}
In this paper, we have explored the role of the sign-changeable
interacting HDE with three infrared (IR) cutoffs, including the
future event horizon, the GO and the Ricci cutoffs in the
framework of BD cosmology. At first, we used the future event
horizon as system's IR cutoff to describe the dynamics of the FRW
Universe by calculating the cosmological parameters such as the
density parameter, the EoS parameter, the deceleration parameter
and the squared sound speed. By choosing  the set of parameters as
$\alpha=10^{-4}$, $\omega=10^4$ and $b^2=0.1$, as the initial
condition, we found out that the density parameter can fill
universe by DE in the long future. We also observed that for $
c\lesssim 1$ the EoS parameter $w_D$ can cross the phantom line
and when $c\geq 1$, we have always  $w_D
>-1$. Besides, for all values of $c$, the deceleration
parameter $q$ transits from deceleration ($q>0$) in the early time
to acceleration ($q< 0$) in the last time. We plotted the
evolution of ${v}^{2}_{s}$ versus $z$ for the different parameters
$\alpha$, $b^2$, $c$ and $\omega$ in Figs.~\ref{S-z1} and
\ref{S-z2}. Graphical analysis of ${v}^{2}_{s}$ shows that in
Fig.~\ref{S-z1} (left panel) for $\alpha <.007$ and in (right
panel) for all values of $b^2$ our model can be stable. Also, in
Fig.~\ref{S-z2} (left panel) by increasing $c$, and in (right
panel) for all of values $\omega$, the sign-changeable interacting
HDE with the future cutoff in BD cosmology can lead to a stable DE
dominated universe.

Furthermore, we have focused on the GO cutoff and observed that by
suitable selection for the model parameters and for all values of
$\beta$, at the late time where the DE dominates we have
$\Omega_D\rightarrow 1$ which have plotted in
Fig.~\ref{OmegaG-z1}. Also, we plotted $w_D $ versus $z$ and see
that in Fig.~\ref{EoSG-z1} (left panel) for $\gamma>1$ we cannot
cross phantom while for $\gamma<1$ for different values of $\beta$
we have $w_D<-1$. The behavior of ${v}^{2}_{s}$ is also plotted in
Figs.~\ref{SG-z1} and \ref{SG-z2}. In Fig.~\ref{SG-z1} we observed
that for different values of $b^2$ (left panel) and $\omega$
(right panel) this model does not allow a stable DE dominated
universe. In Fig.~\ref{SG-z2} (left panel) by taking the
parameters as $\alpha=10^{-4}$, $\omega=10^4$ and $b^2=0.1$ we see
if $\gamma>1$, for different values of $\beta$, ${v}^{2}_{s}$
cannot be stable while for $\gamma<1$ (right panel) (e.g.
$\gamma=0.8$) our model is stable. Finally, we studied these
cosmological parameter when the system's IR cutoff is the Ricci
cutoff in Figs.~\ref{OmegaR-z1} -\ref{SR-z2}. We see three
parameters, $\Omega_D$, $w_D$ and $q$ are consistent with
observations, however, $w_D$ cannot cross the phantom line for
different values of $\omega$. Finally, in Figs.~\ref{SR-z1} and
\ref{SR-z2}, the evolution of ${v}^{2}_{s}$ is plotted. Obviously,
we see that for all different values of $\alpha$, $c^2$, $\omega$
and $b^2$, the squared sound speed ${v}^{2}_{s}$ remains negative
which shows a sign of instability for the HDE in BD theory with
Ricci cutoff.

In conclusion, our studies show that for the sign-changeable HDE
model in the setup of BD cosmology, among the above three IR
cutoffs, the event horizon is the most suitable horizon which can
passes all conditions and leads to a stable DE dominated universe.
%%%%%%%%%%%%%%%%%%%%%%%%%%%%%%%%%%%%%%%%%%%%%%%%%%%%%%%%%%%%%%%%%%%%%%%%%%%%%%%%%%%%%%%
\acknowledgments{We thank Shiraz University Research Council. This
work has been supported financially by Research Institute for
Astronomy \& Astrophysics of Maragha (RIAAM), Iran.}

%%%%%%%%%%%%%%%%%%%%%%%%%%%%%%%%%%%%%%%%%%%%%%%%%%%%%%%%%%%%%%%%%%%%%%%%%%%%%%


\begin{thebibliography}{99}
\bibitem{Riess} A. G. Riess\textit{ et al}, Astron. J. {\bf 116}1009
(1998).

\bibitem{Riess1} S. Perlmutter\textit{ et al}, Astrophys. J.  {\bf 517}  565
(1999).

\bibitem{Riess2} P. deBernardis, \textit{et al},  Nature {\bf 404} 955
(2000).

\bibitem{Riess3} S. Perlmutter,\textit{et al}, Astrophys. J. {\bf 598}102
(2003).

\bibitem{COL2001} M. Colless \textit{et al}, Mon. Not. R. Astron. Soc. \textbf{328}, 1039
(2001).
\bibitem{COL20011} M. Tegmark \textit{etal}, Phys. Rev. D \textbf{69}, 103501
(2004).

\bibitem{COL20012} S. Cole\textit{ et al}, Mon. Not. R. Astron. Soc. \textbf{362}, 505
(2005).
\bibitem{COL20013} V. Springel, C. S. Frenk, and S. M. D. White, , Nature(London) \textbf{440}, 1137
(2006).


\bibitem{HAN2000} S. Hanany\textit{ et al}, Astrophys. J. Lett. \textbf{545}, L5
(2000).

\bibitem{HAN20001} C. B. Netterfield \textit{et al}, Astrophys. J. \textbf{571}, 604
(2002).

\bibitem{HAN20002} D. N. Spergel \textit{et al}, Astrophys. J. Suppl. \textbf{148}, 175
(2003).

\bibitem{Tegmark2004} M. Tegmark et al, Astrophys. J. {\bf606}, 702
(2004).
\bibitem{Tegmark20041} M. Tegmark et al, Phys. Rev. D {bf69}, 103501
(2004).


\bibitem{Ade2014} P.A.R. Ade et al, Astron. Astrophys. {\bf571}, A16
(2014).


\bibitem{Cai2010} R. G. Cai and Q. P. Su, Phys. Rev. D {\bf 81}, 103514
(2010).

\bibitem{Li2004} M. Li,  Phys. Lett. B {\bf603}, 1 (2004).

\bibitem{Zhang2005} X. Zhang, F.Q. Wu,  Phys. Rev. D {\bf72}, 043524
(2005).

\bibitem{Zhang2007} X. Zhang, F.Q. Wu, Phys. Rev. D {\bf76}, 023502
(2007).

\bibitem{Hooft} G. t Hooft,  Conf. Proc. C {\bf930308}, 284 (1993)
\bibitem{Susskind1995} L. Susskind, J. Math. Phys. (N.Y.) {\bf36}, 6377
(1995).

\bibitem{HDE} E. Elizalde, S. Nojiri, S.D.
Odintsov, P. Wang, Phys. Rev. D {\bf 71}, 103504 (2005);\\  B.
Guberina, R. Horvat, H. Stefancic, JCAP {\bf 0505},  001 (2005);\\
B. Guberina, R. Horvat, H. Nikolic, Phys. Lett. B {\bf 636}, 80
(2006);\\ H. Li, Z. K. Guo, Y. Z. Zhang, Int. J. Mod. Phys. D
{\bf15}, 869 (2006);
\\ Q. G. Huang, Y. Gong, JCAP {\bf0408}, 006 (2004) ;\\
J. P. B. Almeida, J. G. Pereira, Phys. Lett. B {\bf636}, 75 (2006)
;
\\  Y. Gong, Phys. Rev. D {\bf70}, 064029 (2004) ; \\ B. Wang, E.
Abdalla, R. K. Su, Phys. Lett. B {\bf611}, 21 (2005);\\
 M. R. Setare, S. Shafei, JCAP {\bf09}, 011 (2006);\\
M. R. Setare, Eur. Phys. J. C {\bf50}, 991 (2007);\\ M. R. Setare,
JCAP
{\bf0701}, 023 (2007) ;\\ M. R. Setare, Phys. Lett. B {\bf654}, 1 (2007) ;\\ M. R. Setare, E. C. Vagenas, Phys. Lett. B {\bf666}, 111 (2008) ;\\
 M. R. Setare, E. N. Saridakis,
Phys. Lett. B {\bf671}, 331 (2009) ;\\I. Duran, L. Parisi ,Phys.
Rev. D {\bf85}, 123538 (2012);
\\F. Yu, Jing-Fei Zhang,  Theor. Phys. {\bf59}, 243 (2013);
\\P. PANKUNNI and TITUS K. MATHEW, Int. J. Mod. Phys. D {\bf23}, 1450024 (2014) ;
\\Y. Hu, M. Li, N. Li, Z. Zhang ,JCAP{\bf08}, 012 (2015);\\HL. Li, JF. Zhang, L.Feng et al. Eur. Phys. J. C {\bf77}, 907 (2017);
\\Ze. Zhao, Shuang. Wang, Sci.China Phys. Mech. Astron. {\bf61},039811
(2018);\\Abdulla Al Mamon, Int. J. Mod. Phys. D {\bf26}, 1750136
(2017);\\Sh. Wang, Yi. Wang, M. Li ,Physics Reports{\bf 696}, 1
 (2017).

\bibitem{Fara} V. Faraoni, (\textit{Cosmology in Scalar-Tensor Gravity}, Kluwer,
Boston, 2004); \\ E. Elizalde, S. Nojiri, S. D. Odintsov, P. Wang,
Phys. Rev. D {\bf71},  103504 (2005); \\ S. Nojiri, S. D.
Odintsov, Gen. Relativ. Gravit. {\bf38},  1285 (2006);\\  R.
Gannouji, et al., JCAP, {\bf0609},  016 (2006).

\bibitem{Brans1961} C. Brans, R.H. Dicke, Phys. Rev. {\bf124}, 925 (1961).
%\textit{Mach's Principle and a Relativistic Theory of Gravitation}%

\bibitem{Bertotti2003}  B. Bertotti,  L. Iess, P. Tortora,
Nature {\bf425}, 374 (2003).

%\textit{A test of general relativity using radio links with the Cassini spacecraft}%
 \bibitem{Felice2006} A.D. Felice,G.Mangano, P.D. Serpico, M. Trodden,  Phys. Rev. D
{\bf74}, 103005 (2006).

\bibitem{Pavon2} N. Banerjee, D. Pavon, Phys. Lett. B {\bf647},  447 (2007).

\bibitem{Setare2} M. R.  Setare, Phys. Lett. B {\bf644}. 99 (2007).

\bibitem{SHBD} A. Sheykhi,  Phys. Lett. B {\bf681}, 205 (2009);\\ A. Sheykhi,
 Phys. Rev. D {\bf81}, 023525 (2010);\\ Ahmad Sheykhi, Mubasher Jamil,  Phys. Lett. B {\bf694}, 284
(2011);\\ E. Ebrahimi, A. Sheykhi,  Phys. Lett. B {\bf706}, 19
(2011);\\ A. Sheykhi, E. Ebrahimi, and Y. Yousefi,  Can.J. Phys.
{\bf91}, 662  (2013).

\bibitem{other} H. Kim, H. W. Lee, Y. S. Myung
 Phys. Lett. B {\bf628},11 (2005);\\ Y. Gong, Phys. Rev. D {\bf70}, 064029 (2004)
;\\ B. Nayak, L. P. Singh, Modern Physics Letters A {\bf24},1785 (2009) ;\\ L. Xu, J.
Lu and W. Li, Eur.Phys.J.C{\bf60},135 (2009);\\ L. Xu, J. Lu  and W. Li, Mod. Phys. Lett. A {\bf25}, 1441 (2010).
\bibitem{Karami} K. Karami, A. Sheykhi, M. Jamil, Z. Azarmi, M. M. Soltanzadeh, Gen. Relativ. Grav.{\bf43},27 (2011);
\\ A. Sheykhi, Phys. Rev. D {\bf81}, 023525 (2010);\\ A. Sheykhi, M. Jamil, Phys. Lett. B {\bf694}, 284 (2011);
\\ A. Pasqua, S. Chattopadhyay, Astrophys. Space Sci. {\bf348}, 284 (2013);\\ V. Fayaz, Astrophys. Space Sci. {\bf361}, 86 (2016);
\\ P. Kumar, C.P. Singh,  Astrophys. Space Sci. {\bf362}, 52 (2017);\\Singh, C.P.  Kumar, P. Int J Theor Phys {\bf56}, 3297 (2017);
\\F. Felegary, F. Darabi, M. R. Setare, Int. J. Mod. Phys. D. {\bf27}, 1850017
(2018).


\bibitem{wang2016} B. Wang, E. Abdalla, F. Atrio-Barandela, D. Pavon, Rep. Prog. Phys., {\bf 79}, 9
(2016).
\bibitem{Wei2011} H. Wei, Nucl. Phys. B {\bf  845}, 381 (2011).
\bibitem{WEI2011} H. Wei, Commun. Theor. Phys.{\bf 56}, 972 (2011).


\bibitem{Signch} M. Khurshudyan, R. Myrzakulov, Eur. Phys. J. C {\bf 77}, 65 (2017);
\\ Ping Xi, Ping Li, Astrophys Space Sci {\bf360}, 3 (2015);\\
Y.D. Xu, Int J Theor Phys,{\bf 52}, 1132 (2013);\\ Y. D.
XU, Com. Theor. Phys. {\bf 65}, 4 (2015);\\
Y.D. Xu and Z.G. Huang, Astrophys Space Sci, {\bf350}, 855 (2014);
\\ M. Abdollahi Zadeh, A. Sheykhi, H. Moradpour, Int. J. Theor.
Phys. {\bf 56}, 3477 (2017); \\Guo, JJ., Zhang, JF., Li, YH. et
al. Sci. China Phys. Mech. Astron. {\bf61}, 030011 (2018).

\bibitem{Peebles20031} P. J. E. Peebles and B. Ratra, Rev. Mod. Phys. {\bf 75} 559 (2003)

\bibitem{StaDE}  Y. S. Myung, Phys. Lett. B {\bf652},  223 (2007);\\
K. Y. Kim, H. W. Lee and Y. S. Myung, Phys. Lett. B {\bf660}, 118 (2008);\\
 E. Ebrahimi and A. Sheykhi, Int. J. Mod. Phys. D {\bf20}, 2369 (2011);\\
E. Ebrahimi and A. Sheykhi, Int. J. Theor. Phys. {\bf52}, 2966
(2013).

\bibitem{Khodam2014} A. Khodam-Mohammadi, E. Karimkhani, and A. Sheykhi,  Int. J. Mod. Phys. D 23, 1450081
(2014).

\bibitem{Golchin2016}E. Ebrahimi, H. Golchin,  Canadian Journal of Physics,{\bf 94}, 1001
(2016).

\bibitem{Arik2006} M. Arik, M.C. Calik, Mod. Phys. Lett. A {\bf 21}1241
(2006).
\bibitem{Arik2008} M. Arik, M.C. Calik and M. B Sheftel, Int. J. Mod. Phys. D {\bf 17}225
(2008).
\bibitem{chimen1} L. P. Chimento,  Phys. Rev. D {\bf81}, 043525 (2010)
\bibitem{chimen2} L. P. Chimento, M. Forte and G. M. Kremer, Gen. Rel. Grav. {\bf41}, 1125
(2009).

\bibitem {Lu2012} J. Lu, L. Ma, M. Liu, Y. Wu, Int. J. Mod. Phys. D {\bf21}, 1250005
(2012).
\bibitem{Granda2008} L. N. Granda, A. Oliveros, Phys. Lett. B {\bf 669}, 275
(2008).
\bibitem{Abdollahi2017} M. Abdollahi Zadeh, A. Sheykhi, H. Moradpour, Int. J. Mod. Phys. D {\bf26}, 8
(2017).

\bibitem{Gao} C. J. Gao, X. L. Chen and Y. G. Shen, Phys. Rev. D {\bf 79}, 043511 (2009).

\end{thebibliography}
\end{document}